\documentclass[12pt]{article}
\usepackage{amsfonts}
\begin{document}

\def\ba{\begin{eqnarray}}
\def\ea{\end{eqnarray}}
\def\a{\alpha}
\def\b{\beta}
\def\k{\kappa}
\def\p{\partial}
\def\o{\omega}
\def\w{\wedge}
\def\f{\frac}
\def\d{\delta}
\def\hb{\hbar}
\def\g{\gamma}
\def\s{\sigma}
\def\ve{\varepsilon}
\def\th{\theta}
\def\vp{\varphi}
\def\pr{\prime}
\def\ua{\uparrow}
\def\da{\downarrow}
\def\Th{\Theta}
\def\D{\Delta}
\def\L{\Lambda}
\def\G{\Gamma}
\def\O{\Omega}

\begin{titlepage}
\title{{\bf Spinor Couplings to Dilaton Gravity
\\ induced by the Dimensional Reduction of \\Topologically Massive
Gravity }}
\author{ M. Adak \footnote{ madak@pamukkale.edu.tr} \\
 {\small Department of Physics, Pamukkale University,}\\
{\small 20100 Denizli, Turkey}  \\
T. Dereli \footnote{ tdereli@ku.edu.tr} \\
 {\small Department of Physics, Ko\c{c} University,}\\
 {\small 34450 Sar{\i}yer, \.{I}stanbul, Turkey  }}
\date{19 January 2004}
\maketitle

\begin{abstract}

\noindent A Dirac spinor is coupled to topologically massive
gravity and  the $D=3$ dimensional action is reduced to $D=2$
dimensions with a metric that includes both the electromagnetic
potential 1-form $A$ and a dilaton scalar $\phi$. The reduced,
electrically charged spinor is made a mass eigenstate with a
(local) chiral rotation. The non-trivial interactions thus induced
are discussed.

\end{abstract}
\end{titlepage}

\section{ Introduction}

Two-dimensional dilaton gravity theories coupled to matter have
attracted a lot of attention because of their relevance to black
holes in effective  string models and also because for some
particular choices of the dilaton and matter couplings they may be
related with higher dimensional theories of gravity. In the
standard Kaluza-Klein theories one usually starts with the
Einstein-Hilbert action for gravity in $D>2$ and obtains a reduced
Lagrangian density in 2-dimensions by inserting the unique
D-dimensional metric-compatible, torsion-free (Levi-Civita)
connection into the original Lagrangian D-form. This connection is
calculated from a special choice of D-dimensional metric that
includes components of vector fields (identified with gauge
potentials) and scalar fields (dilatons) together with the
2-dimensional metric. Many generalizations that involve higher
order curvature invariants in the action and/or connections with
both torsion and non-metricity exist in the literature
\cite{gru},\cite{der3}. We will concentrate our attention here to
$D=3$ dimensions. It is well-known that Einstein's theory in
3-dimensions has no propagating degrees of freedom and no
Newtonian limit.
 A physically interesting modification of the theory is provided
 by the addition of the gravitational Chern-Simons term to the
 action. With this addition, new degrees of freedom are introduced
 and one now has  a dynamical theory that is called topologically
 massive gravity \cite{des1}, \cite{des2}. Recent work on the Kaluza-Klein reduction of
 this theory to $D=2$ dimensions produced an interesting dilaton
 gravity theory and prompted further research \cite{gur},\cite{gru2},\cite{jackiw}.
However, there are not many papers that discuss fermionic matter
couplings to dilaton gravity \cite{cav}. In this paper we couple a
Dirac spinor in $D=3$ dimensions to topologically massive gravity.
The subsequent Kaluza-Klein reduction to $D=2$ dimensions produce
non-trivial interactions that we discuss. We analyze our model in
terms of coordinate independent concepts and make extensive use of
the calculus of exterior forms.

\section{Dimensional reduction  from $D=3$ to $D=2$}

On a 3-dimensional manifold $M_3$ we denote by $\{
\mathbb{X}_A\}$, $A=0,1,2$, an arbitrary 3-frame and the dual
co-frame $\{E^A\}$ by
 \ba
     E^A(\mathbb{X}_B)=\delta^A_B \;.
 \ea
 Frames are declared orthonormal with respect to the general
metric
 \ba
    \mathbb{G} = \eta_{AB} E^A \otimes E^B,
  \ea
where $\eta_{AB} =\mbox{diag}(-,+,+)$. This choice of $\mathbb{G}$
is conserved under local $SO(1,2)$ frame transformations. The
metric-compatible connection 1-forms on $M_3$ satisfy $\Omega_{AB}
= -\Omega_{BA}.$ The torsion and curvature forms associated with
these co-frame and connection forms are given by the structure
equations:
 \ba
     {\mathbb{T}}^A &=& dE^A +{\Omega^A}_B \wedge E^B, \\
     {{\mathbb{R}}^A}_B &=& d{\Omega^A}_B + {\Omega^A}_C \wedge
     {\Omega^C}_B. \label{IR3}
 \ea
The manifold $M_3$ is oriented with the volume 3-form
 \ba
      { }^\# \!1 = E^0\wedge E^1 \wedge E^2
 \ea
where $^\#$  denotes the Hodge map and it is convenient to employ
in the following the graded interior operators
$\imath_{\mathbb{X}_A}$:
 \ba
    \imath_{\mathbb{X}_A} E^B =\delta^B_A \; .
 \ea

Spinors associated with $M_3$ are defined as complex vectors
carrying a representation of the covering group of $SO(1,2)$. To
make contact with the spinors associated with the covering of the
Lorentz group $SO(1,1)$,  we enumerate the Clifford algebra
generators $\Gamma_A$ associated with the orthonormal frames of
$M_3$ satisfying
 \ba
     \{\Gamma_A , \Gamma_B \} = 2 \eta_{AB}
 \ea
with the choice
 \ba
   \Gamma_0  &=&  \left (
                 \begin{array}{cc}
                                   0   & 1  \\
                                   -1  & 0
                 \end{array}
          \right )             \; , \;\;\;
   \Gamma_1  =   \left (
                 \begin{array}{cc}
                                   0    &  1  \\
                                   1 &  0
                 \end{array}
          \right )  \; , \;\;\;
 \Gamma_2  =   \left (
                 \begin{array}{cc}
                                   1 &  0  \\
                                   0 &  -1
                 \end{array}
          \right )  \; .            \label{dirmat}
 \ea
The generators of the Clifford algebra associated with the
orthonormal frames of space-time $M_2$ are
 \ba
       \{\gamma_a , \gamma_b \} = 2 \eta_{ab} \quad , \quad a=0,1
 \ea
where $\eta_{ab} =\mbox{diag}(-,+)$. The algebra of $SO(1,2)$ is
generated by
 \ba
     \Sigma_{AB} = \frac{1}{4}[\Gamma_A , \Gamma_B ]
 \ea
and that of $SO(1,1)$ by
 \ba
     \sigma_{ab} = \frac{1}{4}[\gamma_a , \gamma_b ]\; .
 \ea
 A 2-component complex spinor $\Psi$ in $D=3$ dimensions is constructed
locally from a set of two complex functions on $M_3$. It will be
reduced in a certain way below to a 2-component complex spinor
$\psi$ on $M_2$.

Three local co-ordinate functions for $M_3$ will be labelled by
$x^A \equiv (x^a,y)$. The submanifold $y=0$ in this chart is then
$M_2$ which inherits a patch with two co-ordinates $(x^0,x^1)$.
Now we write Kaluza-Klein metric as
 \ba
     \mathbb{G}= g + \phi^2 A \otimes A +\phi^2( dy \otimes
     A + A \otimes dy) + \phi^2 dy \otimes dy
 \ea
where the metric on $M_2$ is
 \ba
     g = -e^0 \otimes e^0 +  e^1 \otimes e^1 \; .
   \ea
A particularly convenient choice of the orthonormal 3-frame  is
 \ba
     {\mathbb{X}}_a &=& X_a -(\imath_{X_a}A)\partial_y \; , \\
     {\mathbb{X}}_2 &=& \frac{1}{\phi (x)} \partial_y \; ,
 \ea
where $\partial_y =\frac{\partial}{\partial y} \in T(M_3)$ is a
vector field, $\phi (x)$ and $A(x)$ are respectively $0$ and $1$
forms on $M_2$ and $X_a \in T(M_2)$. The dual co-frame is
 \ba
      E^a &=& e^a(x) \; , \\
      E^2 &=& \phi (x) [dy +A(x)]\; ,
 \ea
where $e^a(X_b)= \delta^a_b$ and $e^a \in \Lambda^1(M_2)$ satisfy
the structure equations
\ba
  de^a +{\omega^a}_b \wedge e^b =  0 \;, \;
  {R^a}_b =  d{\omega^a}_b \; .
 \ea
on $M_2$. The orientation of $M_2$ is given by the volume element
 \ba
     ^*1 = e^0 \wedge e^1
 \ea
where ${ }^*$  is the Hodge map. The field $\phi (x)$ simply
scales the vector field $\partial_y$ to normalize it with respect
to the metric $\mathbb{G}$. Using this reduction scheme we can
reduce the Levi-Civita connection 1-forms ${\Omega^A}_B$ as
follows:
 \ba
   {\Omega^0}_1 &=& {\omega^0}_1 + \frac{\phi}{2} f E^2 \; , \nonumber \\
   {\Omega^0}_2 &=&   \frac{\phi}{2} f e^1 - \frac{\partial^0\phi}{\phi}E^2 \; , \nonumber \\
   {\Omega^1}_2 &=&  \frac{\phi}{2} f e^0 - \frac{\partial^1\phi}{\phi}E^2
    \label{Omega}
 \ea
 where
 \ba
      F = dA = f *1 \;.
\ea
 The corresponding curvature 2-forms are
 \ba
   {{\mathbb{R}}^0}_1 &=& {R^0}_1  + \frac{3}{4}\phi^2 f^2 { }^*1
   +(\frac{3}{2} f d\phi + \frac{\phi}{2} df ) \wedge E^2 \; , \nonumber \\
    {{\mathbb{R}}^0}_2 &=& \frac{1}{2}d(\phi f) \wedge e^1 - \partial^0\phi F
     -[ \frac{1}{\phi} D(\omega )(\partial^0\phi ) + \frac{\phi^2}{4} f^2 e^0 ]\wedge E^2
     \; , \nonumber \\
    {{\mathbb{R}}^1}_2 &=& \frac{1}{2}d(\phi f ) \wedge e^0  - \partial^1\phi F
     -[ \frac{1}{\phi}D(\omega )(\partial^1\phi ) + \frac{\phi^2}{4} f^2 e^1 ]\wedge E^2
     \;. \label{R3}
 \ea

We consider the action $I[E, \Omega ] = \int_{M_3} \mathbb{L}$
where the Lagrangian 3-form
 \ba
     \mathbb{L} = \kappa {\mathbb{R}^A}_B \wedge ^\#(E_A \wedge E^B) +
     \frac{\mu}{2} ({\Omega^A}_B \wedge  d{\Omega^B}_A
     +\frac{2}{3}{\Omega^A}_C \wedge  {\Omega^C}_B \wedge  {\Omega^B}_A )
     + \lambda ^\#1  \label{L3}
 \ea
contains the Einstein-Hilbert term with the coupling constant
$\kappa$,  the gravitational Chern-Simons term with the coupling
constant $\mu$ written in terms of Levi-Civita connection 1-forms
${\Omega^A}_B$ and a cosmological constant $\lambda$. A study of
this theory can be found in Ref.\cite{der2}. We dimensionally
reduce (\ref{L3}) and obtain a reduced Lagrangian 3-form
$\mathbb{L} = L \wedge dy$ where
 \ba
   L &=& \kappa [\frac{\phi}{2}\mathcal{R} + \frac{\phi^3}{4} f^2 ]{ }^*1
          +\phi \lambda ^*1  \nonumber \\
     & &  + \mu [  (\frac{\phi^2}{2} f \mathcal{R}  + \frac{\phi^4}{2} f^3 ){ }^*1
          - 2 f d\phi \wedge ^*d\phi  - \phi df \wedge ^*d\phi ] \;.
 \ea
that agrees with \cite{gur}, \cite{gru2} when $\phi$ is a
constant. $\mathcal{R}$ is the scalar curvature on $M_2$.

We now consider the  Dirac Lagrangian 3-form
 \ba
    \mathbb{L}_D = \mbox{Her} (i\overline{\Psi} \; { }^\# \Gamma \wedge \mathbb{D} \Psi )
     + iM\overline{\Psi} \Psi \,{ }^\#\!1 \label{dirac3}
 \ea
in terms of the Clifford algebra ${\mathcal{C}}\ell_{1,2}$-valued
1-forms $ \Gamma = \Gamma_A E^A $ and  $M=\frac{mc}{\hbar}$ is the
Dirac mass. The covariant exterior derivative  of a Dirac spinor
is given explicitly as
 \ba
  \mathbb{D}\Psi = d\Psi + \frac{1}{2}\Omega^{AB}\Sigma_{AB} \Psi \; .\label{covder}
 \ea
Our reduction ansatz \cite{der1} for the spinors is
 \ba
     \Psi (x,y) &=& e^{\gamma_5 \lambda (x)+iqy} \psi (x) \;,\\
     \overline{\Psi} (x,y) &=& \Psi^+ \Gamma_0 = \overline{\psi} e^{\gamma_5 \lambda (x)-iqy}
 \ea
where $\lambda (x) $ is a real function on $M_2$ to be determined
later, $q$ is a constant identified with the electric charge and
 \ba
     \Gamma_a = \gamma_a \quad , \quad \gamma_5 = i\gamma_0 \gamma_1 =
     i\Gamma_2 \; .
 \ea
Here we assumed that the spinor $\Psi$ is periodic in $y$ in order
to obtain a minimal coupling to electromagnetism governed by the
electromagnetic 1-form $A$. Now we first calculate
 \ba
      ^\#\Gamma = - i\gamma_5 ^*1 - E^2 \wedge ^*\gamma
 \ea
where $\gamma = \gamma_a e^a$ is the Clifford algebra
${\mathcal{C}}\ell_{1,1}$-valued 1-form and then
 \ba
     \mathbb{D} \Psi = e^{\gamma_5 \lambda (x)+iqy} [D\psi + \gamma_5 d\lambda (x) \psi
                        +E^2 i( \frac{q}{\phi} - \frac{\phi}{4}f \gamma_5)\psi ]\nonumber \\
         +e^{-\gamma_5 \lambda (x)+iqy}[  i \frac{\phi}{4} f { }^*\gamma \gamma_5
         + E^2 i \frac{1}{2\phi} { }^*(^*\gamma \wedge d\phi
         )\gamma_5]\psi \;. \label{31}
 \ea
Here
 \ba
     D\psi = d\psi +\frac{1}{2}\omega^{ab}\sigma_{ab}\psi -iqA
     \psi \; . \label{Dpsi}
 \ea
is the $U(1)$-covariant exterior derivative of a spinor and
physically describes the minimal interactions with gravity and
electromagnetism. The second term on the right hand side of
(\ref{31}) simulates a chiral interaction that senses the
handedness of the $\psi$ spinor if $\lambda$ is not a constant.
With these definitions, the dimensionally reduced Dirac Lagrangian
2-form becomes
 \ba
     L_D = \phi \{ -\mbox{Her} (i\overline{\psi} \; { }^*\gamma \wedge D\psi )
     + i\overline{\psi}\; e^{2\lambda \gamma_5} (\frac{q}{\phi}\gamma_5
     +M)\psi ^*1 + i d\lambda \wedge \overline{\psi} \; { }^*\gamma \gamma_5 \psi \nonumber \\
     +\frac{i}{2\phi} \overline{\psi} \; d\phi \wedge { }^*\gamma   \psi
     -i \frac{\phi}{4} f \overline{\psi} \; e^{2\lambda \gamma_5} \psi { }^*1 \}  \; . \label{dirac2}
 \ea
The first term in (\ref{dirac2}) is the kinetic term, the second
is the mass term and the rest are new interaction terms. However,
$\gamma_5$ in the mass term means that the left and right handed
parts of the reduced 2-component  spinor would have different
masses. Since we wish to identify both the left and the right
handed parts of the reduced 2-component spinor with the chiral
components of a Dirac particle of definite mass, we set
 \ba
     \tan{2\lambda} = -\frac{q}{M\phi} \; . \label{lambda}
 \ea
Here since ${\gamma_5}^2 =-1$ we write
 \ba
    e^{2\lambda \gamma_5} = \cos{2\lambda} +\gamma_5 \sin{2\lambda} \; .
 \ea
It follows from (\ref{lambda}) that
 \ba
    d\lambda = \frac{qM}{2(q^2 +M^2\phi^2)} d\phi \;.
 \ea
Thus we write down the reduced Dirac Lagrangian 2-form as
 \ba
   L_D &=& \phi \{ -\mbox{Her} (i\overline{\psi} \; { }^*\gamma \wedge D\psi )
   -i \mathcal{M} \overline{\psi} \; \psi ^*1 \nonumber \\
 & &  + i \frac{1}{2\phi}d\phi \wedge \,\overline{\psi} \;  { }^*\gamma \psi
  + i \frac{qM}{2\mathcal{M}^2\phi^2}d\phi \wedge \, \overline{\psi}  \;  { }^*\gamma \gamma_5 \psi
  \nonumber \\
 & & + i\frac{f M \phi}{4 \mathcal{M}} \overline{\psi}\; \psi { }^*1
 - i\frac{q f}{4 \mathcal{M}} \overline{\psi}\; \gamma_5 \psi { }^*1
   \} \label{decom-dirac}
 \ea
where the effective mass is
 \ba
     \mathcal{M} = \sqrt{M^2 + \left( \frac{q}{\phi} \right)^2} \; .
 \ea

\section{Conclusion}

We have coupled a Dirac spinor $\Psi$ to topologically massive
gravity and dimensionally reduced the $D=3$ dimensional action to
$D=2$ dimensions with a Kaluza-Klein metric that includes both the
electromagnetic potential 1-form $A$ and a dilaton scalar $\phi$.
The reduced, electrically charged spinor $\psi$ becomes a mass
eigenstate after a (local) chiral rotation with (\ref{lambda}).
The Lagrangian 2-form (\ref{decom-dirac}) summarizes all the
non-trivial interactions thus induced. In particular the effective
mass $\mathcal{M}$ depends on the 3-dimensional mass $M$, the
electric charge $q$ as well as the dilaton field $\phi$. Even if
we initially set $M$ to zero in 3-dimensions, there may still be a
non-zero effective mass induced in 2-dimensions with $\lambda = -
\frac{\pi}{4}$ provided $q \neq 0$. There are interactions that
describe vector and pseudo-vector couplings to the gradient of the
dilaton field. Finally, the last two terms in (\ref{decom-dirac})
describe non-minimal electric and magnetic dipole moment
interactions, respectively. They provide additional contributions
to the established gyromagnetic ratio defined by the minimal
electromagnetic coupling to the spinor. It is interesting to note
that both the dipole moments have explicit dependence on the
dilaton field. We intend to investigate solutions to the induced
Dirac equation in a separate paper.

\vskip 2cm

\noindent {\large {\bf Acknowledgement}}

\noindent One of the authors (MA)  is supported by the Scientific
Research Project (BAP) 2002FEF007, Pamukkale University, Denizli,
Turkey.

\end{document}